\documentclass[twocolumn,showpacs,preprintnumbers,amsmath,amssymb]{revtex4}
\usepackage{graphicx}% Include figure files
\usepackage{dcolumn}% Align table columns on decimal point
\usepackage{bm}% bold math

\begin{document}

\title{Interplay between Superconducting and Pseudogap States Revealed by
Heating-Compensated Interlayer Tunneling Spectroscopy on
Bi$_2$Sr$_2$CaCu$_2$O$_{8+x}$ }

\author{Myung-Ho Bae$^1$}
\author{Jae-Hyun Choi$^1$}
\author{ Hu-Jong Lee$^1$}
\author{Kee-Su Park$^2$}
\affiliation{$^1$Department of Physics, Pohang University of
Science and Technology, Pohang 790-784, Republic of Korea}%
\affiliation{$^2$Department of Chemistry, Pohang University of
Science and Technology, Pohang 790-784, Republic of Korea}%

\date{\today}

\begin{abstract}
The heating-compensated interlayer tunneling spectroscopy is
performed on stacks of Bi$_2$Sr$_2$CaCu$_2$O$_{8+x}$ intrinsic
junctions. The high-accuracy spectra without the local heating,
for varying temperatures and magnetic fields, reveal that the
spectral weight forming the superconducting coherence peak is
mainly contributed from the dip position of the hump structure of
the pseudogap state. The observed U-shaped subgap structure is
consistent with the weighted antinodal tunneling between Cu-O
double layers in a junction as suggested in Ref. \cite {hop}.
\end{abstract}


\maketitle

Since the discovery of high-$T_c$ superconductivity in cuprates,
various anomalous behaviors have been discovered beyond the
standard BCS theory. One of them is the pseudogap (PG) observed in
the normal state of the materials, which has aroused a great deal
of controversy about the relationship between the PG and the
superconducting state \cite{review}. The dispute has been focused
on two points. One scenario argues that the PG is a precursor to
the superconducting state. The other, however, considers it only
as a competing or coexisting order with the superconductivity.

Recently, a variety of powerful measurement tools are developed to
study the PG state as well as the superconducting state. The
surface probes like the angle-resolved photoemission spectroscopy
(ARPES) and the scanning tunneling spectroscopy (STS) on
Bi$_2$Sr$_2$CaCu$_2$O$_{8+x}$ (Bi-2212) single crystals have
suggested that the PG is the paired precursor state of the
superconducting coherence \cite{precursor}. More recent STS
studies, however, have shown that the PG is a competing or
coexisting order with the superconductivity on the same Fermi
surface \cite{STS}. The interlayer tunneling spectroscopy (ITS),
which probes the bulk tunneling properties of intrinsic junctions
\cite {Kleiner} imbedded in single crystals, also seems to reveal
that the two states are separate in their physical origins, as
manifested by the peak-dip-hump (PDH) structure below the
superconducting critical temperature $T_c$ \cite{ITS}. The PDH
characteristics in the ITS used to be considered as the
superposition of the superconducting coherence peak (CP) and the
PG hump structure below $T_c$.

The onset of the superconductivity in cuprates is accompanied by
the emergence of a sharp CP that forms the gap edge on the
background of pre-depleted quasiparticle DOS near the Fermi
surface. Clearer distinction between the CP and the hump reveals
in an external magnetic ($H$) field applied along the $c$ axis. An
$H$ field, a tuning parameter of the phase coherence, reduces the
spectral weight in the CP and redistributes it in a fashion
corresponding to the state of lower condensation. As in the BCS
case \cite{Levine}, the total spectral weight is expected to be
conserved in the process. Existing ITS measurements in Bi-2212
near $T_c$, however, revealed that, as the CP reduces
significantly in $H$ fields, the spectral weight in the subgap
(the bias region below the CP) remains almost unaltered or even
decreases \cite{mag1,mag2}. Since, in these studies, the spectral
weight of the hump was not altered in the process, either, the
results imply that the total spectral weight in varying fields is
not conserved, which is hard to be accepted. The delicate transfer
of the relative spectral weight between the CP and the hump, which
can be a key to understanding the basic mechanism of the
high-$T_c$ superconductivity, is thus yet to be clarified
accurately.

In this letter we report the detailed PG behavior both in the
normal and the superconducting states and analyze the interplay
between the PG and the superconducting states, on the basis of the
temperature ($T$), the magnetic field ($H$), and the doping ($p$)
dependencies of the spectrum. Recently, it has been found that the
ITS is susceptible to the self-heating caused by a finite bias in
combination with the poor thermal conductivity of the Bi-2212
\cite{heat}. To eliminate the self-heating, we adopted the
heating-compensated constant-$T$ ITS on stacks of both overdoped
and underdoped Bi-2212 intrinsic junctions \cite{Bae2}. Using the
{\it bona-fide} electronic spectra from our improved ITS the
transfer of the spectral weight was traced with high accuracy in
varying $H$ fields, which leads to clarifying the origin of the
PDH structure and the relation between the PG and the
superconductivity. In addition, the pronounced U-shaped subgap
spectra observed in our ITS constitute another distinctive feature
from the existing ITS results, providing an insight into the
nature of the interlayer tunneling.

\begin{figure}[b]
\begin{center}
\leavevmode
%h=here, t=top, b=bottom, p=separate figure page
\includegraphics[width=0.65\linewidth]{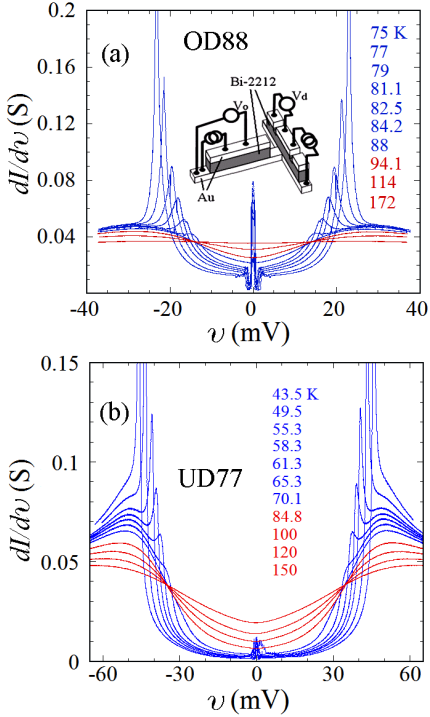}
\caption{(color online) The interlayer tunneling spectra $dI/dv$
for (a) the overdoped OD88 and (b) the underdoped UD77 as a
function of bias voltage per junction at various $T$. Inset of
(a): the measurement configuration.}
\end{center}
\end{figure}

The Bi-2212 single crystals were prepared by the
solid-state-reaction method (UD75, OD88) and the
traveling-solvent-floating-zone method (UD77) \cite{Oda}. Instead
of a usual mesa structure, a sample was prepared into a 3$\times$3
$\mu$m$^2$ stack of intrinsic junctions sandwiched between two
Au-film electrodes, by employing the double-side-cleaving of
Bi-2212 crystals, photo-lithographic micropatterning, and ion-beam
etching \cite{Bae}. A sample consisted of two stacks [the inset of
Fig. 1(a)]; the left one represents the sample stack on which the
ITS was taken. The right one, placed a fraction of micrometer
apart from the sample stack and consequently in strong inter-stack
thermal coupling through the common bottom Au electrode, is for
the {\it in-situ} thermometry of the sample stack. The
proportional-integral-derivative control of the constant sample
temperature during the ITS in a finite bias is described in detail
in Ref. \cite{Bae2}. The temperature during a sweep was kept
constant within $\sim$0.1 K for a bias per junction up to
$\sim$120 mV ($\sim$70 mV, $\sim$40 mV) for UD75 (UD77, OD88).
Measurements were made along the highest-bias quasiparticle curve,
{\it i.e.}, the last branch. The $dI/dV$ curves were obtained
using the lock-in technique operating at 33.3 Hz.

The $c$-axis superconducting transition temperatures $T_c$ were
75.2 K (UD75), 77.0 K (UD77), and 88.3 K (OD88). The corresponding
doping levels determined by the empirical relation \cite{ITS},
$T_c$=95[1-82.6($p$-0.16)$^2$], were $p$=0.109, 0.112, and 0.19,
respectively. The underdoped(UD75 and UD77) and overdoped (OD88)
sample stacks contained $N$=15, 24, and 19 intrinsic junctions,
respectively, as determined by the number of quasiparticle
branches in the zero-field {\it I-V} curves at 4.2 K (not shown).

A series of interlayer tunneling spectra $dI/dv(v)$ of OD88 and
UD77, for varying $T$, are displayed in Fig. 1, where the voltage
is normalized by the number of junctions as $v$=$V/N$. In the
normal state of both samples above $T_c$ the low-bias DOS is
smoothly depleted, revealing the PG. At a $T$ below $T_c$ a
sharper peak (the CP) develops inside the PG, constituting the PDH
structure. Further lowering $T$, the fast sharpening CP with the
growing SG size overwhelms the spectrum, leaving only the CP
apparent. The PG with the hump is more conspicuous in the
underdoped UD77. Both samples below $T_c$ show the U-shaped DOS in
the subgap region, which is contrasted to the mainly V-shaped DOS
observed previously \cite{ITS}. An interesting observation from
all our samples is that all the ITS curves in the PG state above
$T_c$ intersect at a single point inside the PG. Similar feature
was observed in the ARPES measurements on the antinodal region of
the Fermi surface above $T_c$ \cite{Norman}, which may provide an
evidence that the interlayer tunneling is dominated by the
antinodal hopping. The fluctuating conductance at zero bias
sufficiently below $T_c$ in Fig. 1 was caused by the Josephson
pair tunneling.

%The normal state of OD88 in the inset of Fig. 1(a) shows distinct
%zero-bias depletion of DOS with the PG size for each $T$ denoted
%by a pair of vertical segments. The DOS depletion around zero bias
%persisted up to the maximum $T$ examined, 241 K. Since the
%zero-bias tunneling resistance is expected to increase upon
%opening of the PG the $T$ for the minimum tunneling resistance
%($T_{min-R}$) used to be conveniently assigned as the PG onset
%temperature, $T^*$. In contrast to this expectation and existing
%observations \cite{Kawakami}, however, the depletion of the DOS
%near zero bias is evident in our constant-$T$ ITS even above
%$T_{min-R}$ ($\sim$170 K in OD88), which is in clear contradiction
%to $T^*$ ($>$241 K) where the dip is supposed to disappear. As
%pointed out recently \cite{Yurgens}, the erroneous assignment of
%$T^*$ was caused by the self-heating. For UD77 and UD75 the PG
%persisted even at room temperature.

\begin{figure}[b]
\begin{center}
\leavevmode
%h=here, t=top, b=bottom, p=separate figure page
\includegraphics[width=1\linewidth]{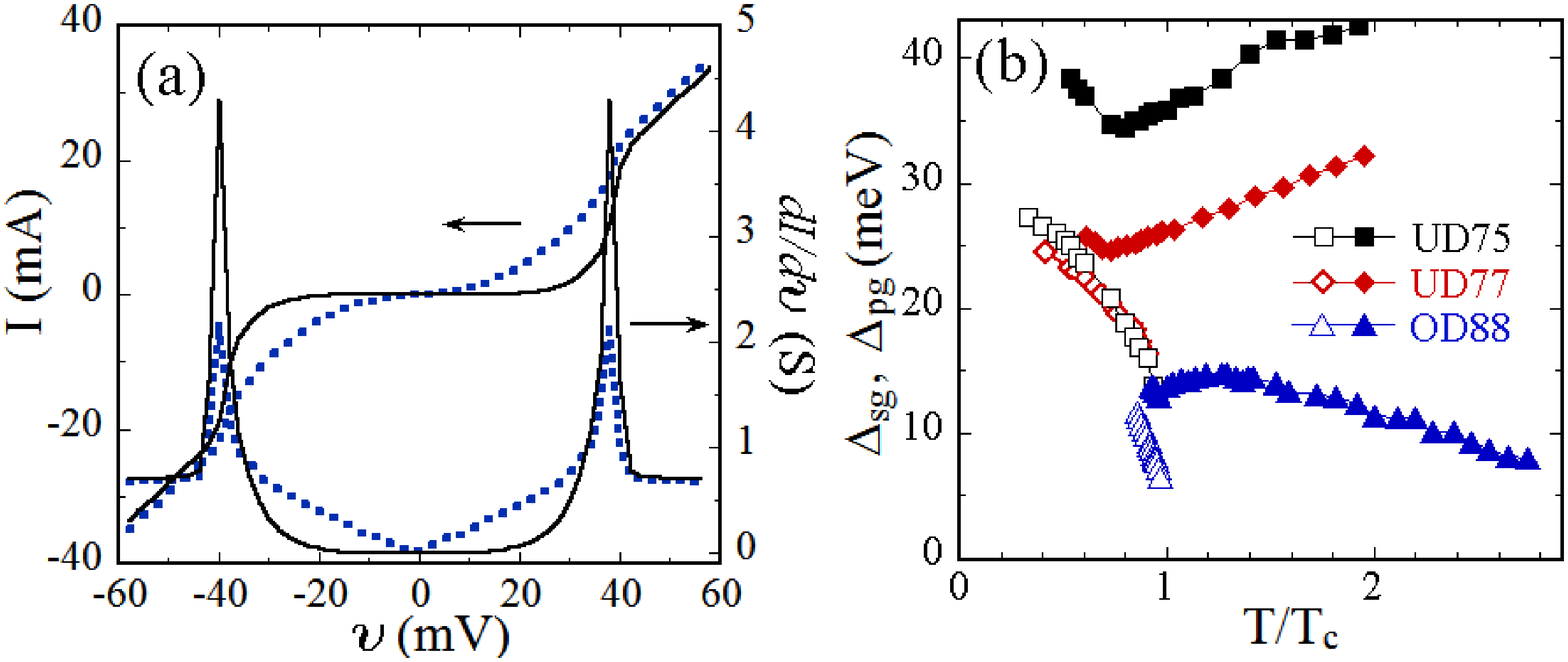}
\caption{(color online) (a) Calculated {\it I-V} and $dI/dv(v)$
curves with Eq. (1) for $T_{\phi}$=1 (dotted curves) and
$T_{\phi}$=$t_{\bot}\mbox{cos}^2 2\phi$ (solid curves) at $T$=4.2
K. (b) The $T$ dependencies of the SG ($\Delta_{sg}$, open
symbols) and the PG ($\Delta_{pg}$, filled symbols).}
\end{center}
\end{figure}

The U-shaped DOS in the subgap region poses an important
implication to the interlayer tunneling in high-$T_c$
superconductors. The tunneling quasiparticle current in a
Josephson junction is written as $I(V)=4\pi e
\sum_{\vec{k}\vec{p}}|T_{\vec{k}\vec{p}}|^2[f(\xi_{\vec{k}})
-f(\xi_{\vec{p}})]\delta(eV+\xi_{\vec{k}}-\xi_{\vec{p}})$, where
$T_{\vec{k},\vec{p}}$ is the tunneling matrix element,
$\xi_k\equiv\sqrt{E_k^2-\Delta_k^2}$, and $f(\xi_k)$ is the Fermi
function \cite{Mahan}. Assumed coherent elastic tunneling
$|T_{\vec{k},\vec{p}}|^2=|T_{\vec{k}}|^2\delta_{\vec{k},\vec{p}}$
\cite{coherent}, along with the $d_{x^2-y^2}$-wave gap symmetry
and the relation $\sum_{\vec{k}}=\frac{A}{(2\pi)^2}\int
d\vec{k}\int d\xi_{\vec{k}}$, leads to
\begin{eqnarray}
I(V)&=&\frac{1}{2\pi eR_n}\int_0^{2\pi}d\phi |T_{\phi}|^2\int
 dE N(\phi,E)N(\phi,E+eV)\nonumber \\
& & \times\{f(E)-f(E+eV)\},
\end{eqnarray}
where $A$ is the junction area, $\phi \equiv$ tan$^{-1}(k_y/k_x)$,
and $N(\phi,E)=\mbox{Re}\{(E-i\Gamma)/[(E-i\Gamma)^2-
\Delta_0^2\mbox{cos}^22\phi]^{1/2}\}$ with the quasiparticle
scattering rate, $\Gamma$. The dotted curves in Fig. 2(a) are the
numerical result for a $k$-independent isotropic tunneling matrix
element $T_{\phi}$=1 \cite{Yamada} with $\Delta_0$=20 meV and
$\Gamma$= 0.05 meV at $T$=4.2 K. On the other hand, the band
calculation for a crystal with the tetragonal symmetry predicts
that the $c$-axis quasiparticle tunneling dominates near the
antinodal points in the first Brillouin zone, satisfying
$T_{\phi}=t_{\bot}\mbox{cos}^2 2\phi$ \cite{hop}. The solid curves
in Fig. 2(a) correspond to this anisotropic $T_{\phi}$, which
reduces the low-energy quasiparticle tunneling near the nodal
points. It leads to the U-shaped tunneling conductance, while
sharpening the CP \cite{Su}. Thus, the pronounced U-shaped spectra
in our ITS (Fig. 1) strongly suggest again that the antinodal
hopping is highly weighted in the interlayer tunneling.

The $T$ dependence of the SG energy $\Delta_{sg}$ (open symbols)
and the PG energy $\Delta_{pg}$ (filled symbols) is shown in Fig.
2(b), which clearly reveals the coexistence of the CP and the
hump. For each doping level $\Delta_{sg}$ reduces as $T$
approaches $T_c$. By contrast, the $\Delta_{pg}$ exhibits a more
complicated $T$ dependence. For UD77 and UD75 in the normal state
above $T_c$, $\Delta_{pg}$ shrinks with decreasing $T$. But
overdoped OD88 shows a local maximum of $\Delta_{pg}$ above $T_c$.
The $\Delta_{pg}$ in each doping level reaches a local minimum at
a $T$ slightly below the respective $T_c$ and rises with further
lowering $T$. In OD88, $\Delta_{pg}$ was observed to grow from
zero with lowering $T$ below $T^*$. UD77 and UD75 are expected to
exhibit similar behavior, except for more extended temperature
scales.

We now focus on whether the dip is caused by the simple
superposition of the CP and the PG hump or by the transfer of the
spectral weight. As in OD88 and UD77, the dip in the
superconducting state in our ITS disappears when $\Delta_{sg}$
becomes comparable to $\Delta_{pg}$ with lowering $T$ below $T_c$.
For UD75, however, the PDH structure exists even down to $\sim$20
K because $\Delta_{pg}$ is sufficiently larger than $\Delta_{sg}$
at the temperature. Fig. 3 illustrates the progressive evolution
of the simple hump structure into the PDH one in UD75 with
decreasing $T$. The CP grows sharply with decreasing $T$ below
$T_c$. Along with it, the position of the dip shifts to a higher
voltage with deepening its depth. In the PDH structure the DOS in
the hump drops suddenly at the dip position while sharpening the
CP, which becomes more distinct at lower $T$.

\begin{figure}[b]
\begin{center}
\leavevmode
%h=here, t=top, b=bottom, p=separate figure page
\includegraphics[width=0.65\linewidth]{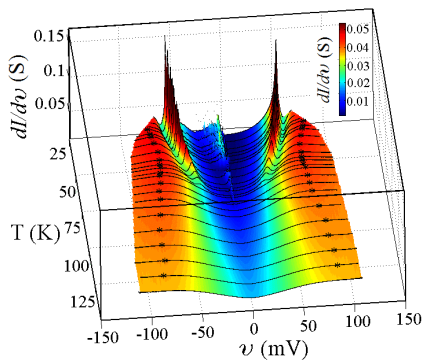}
\caption{(color) Color-coded plot of the interlayer tunneling
spectra per junction for UD75 for varying $T$. The solid curves
are the measured spectra, with the edge of $\Delta_{pg}$ denoted
by * symbols.}
\end{center}
\end{figure}

The redistribution of the spectral weight is more clearly visible
in a high $H$ field, applied in this study up to 6 T along the $c$
axis. As in the case of varying $T$, for a varying phase coherence
in an $H$ field, the relative weight of the DOS redistributes
between the CP and the hump. Figs. 4(c) and (d) [Figs. 4(g) and
(h)] display the $H$-field dependence of the spectra for UD75
[OD88] at 65 and 55 K [82.6 and 80 K], respectively. The insets of
Figs. 4(d) and (g) [Fig. 4(e)] show the total spectral weight,
obtained by integrating the respective curve for each $H$ field
[temperature] as in the main panels between two cutoff voltages
(-$v_c$, $v_c$) and normalizing it by the corresponding zero-field
[79 K] value. The cutoff voltage $v_c$ of integration was selected
well above the merging position of the different-$H$ and
different-$T$ spectra around the hump \cite {Integ}. The resulting
virtually $H$- and $T$-independent total spectral weight is
consistent with the sum rule \cite {Randeria} and, thus, confirms
the reliability of our description for the transfer of the
spectral weight based on our ITS.

%Unlike the other two samples, the DOS of UD75 redistributes more
%conspicuously for varying $H$ fields.

\begin{figure}[t]
\begin{center}
\leavevmode
%h=here, t=top, b=bottom, p=separate figure page
\includegraphics[width=1\linewidth]{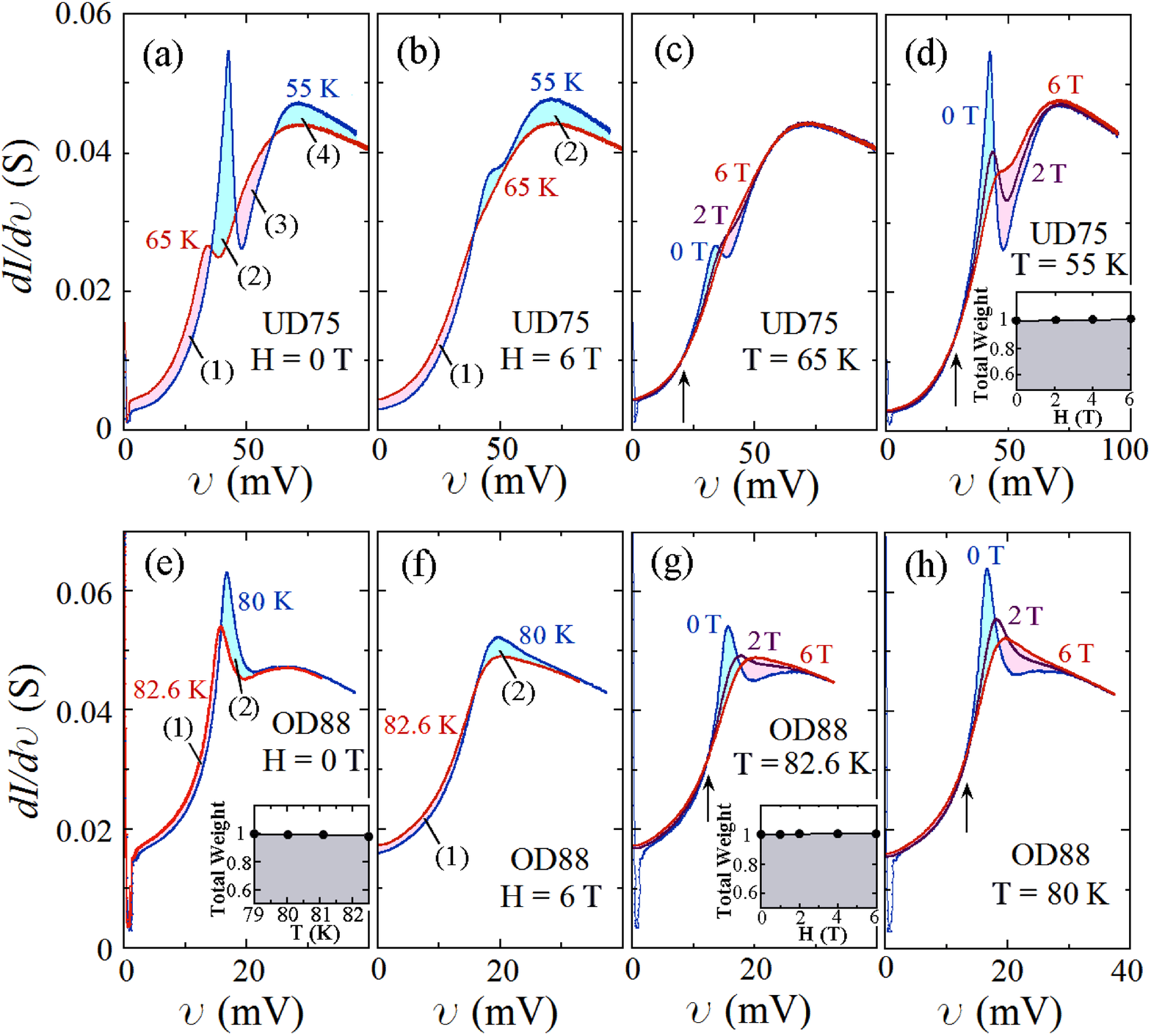}
\caption{(color) (a) and (e): zero-field spectrum change for sets
of temperatures for UD75 and OD88. (b) and (f): the corresponding
curves in $H$=6 T, almost representing the PG structure at given
$T$ for the two samples. (c), (d), (g), and (h): $H$-field
dependence of the peak-dip-hump structure for the corresponding
$T$. Insets: the total spectral weight vs $T$ [(e)] and $H$ field
[(d) and (g)].}
\end{center}
\end{figure}

The heat-compensated constant-$T$ ITS enabled the accurate tracing
of the spectral weight transfer by any external parameters
changing the coherence, such as $T$ and $H$ \cite {homogeneity}.
The spectra of UD75 in Fig. 4(d), at a fixed $T$ of 55 K, reveal
that the CP is suppressed rapidly while the dip fills up with
increasing $H$ field. Similar trend holds at 65 K [Fig. 4(c)],
although the CP and the corresponding spectral change are much
less pronounced. This trend also holds for the overdoped OD88 with
a smeared PDH structure as in Figs. 4(h) and (g). In comparison,
in the process, only a small portion of the CP spectral weight is
transferred to the subgap region, especially for UD75. In
conventional BCS superconductors the spectral weight composing the
CP would be all transferred from the subgap region. In high-$T_c$
superconductors, however, large depletion of the spectral weight
in the Fermi surface already exists in the normal state below
$T^*$ \cite {review}. Thus, upon onset of the superconductivity
the condensation of the quasiparticles is limited to the Fermi arc
in the vicinity of ($\frac{\pi}{2}, \frac{\pi}{2}$) point
\cite{Norman}, corresponding to the low-bias subgap region in the
tunneling spectra, which makes a relatively minor contribution to
the CP. Recent ARPES measurements indicate that the Fermi arc
extends for higher doping \cite{arc}, allowing more quasiparticles
to condense to form the CP along with. That may explain why in
Fig. 4, as the superconducting coherence is enhanced, a higher
portion of quasiparticle spectral weight below the crossing points
(denoted by the arrows) is transferred to the CP in OD88 than in
UD75.

The $T$ dependence of the spectral shape in Figs. 1(a) and (b) as
well as Fig. 3 appears to suggest that rapid growing of the CP
with decreasing $T$ below $T_c$ is caused by the fast depletion of
the spectral weight below the gap as the superconductivity sets
in. The trend is more clearly seen in the comparative illustration
of the zero-field DOS for a couple of temperatures below $T_c$ for
UD75 [Fig. 4(a)] and OD88 [Fig. 4(e)]. For UD75, as $T$ decreases
from 65 K to 55 K, the spectral weight transferred below the gap
[Region (1)] and in the dip [Region (3)] constitutes the CP
[Region (2)] and the hump [Region (4)]. For OD88, similar transfer
of spectral weight takes place from Regions (1) to (2) as $T$
lowers from 82.6 K to 80 K. Although the superconductivity is not
completely suppressed in 6 T [as evidenced by the presence of the
shoulder near 40 mV of Fig. 4(b)] the spectral curves at different
$T$ in Fig. 4(b) and (f) almost represent the PG structure at the
respective temperatures \cite {PG structure}. This indicates that
the apparent further depletion of the low-energy spectral weight
with lowering $T$ [Region (1) of Figs. 4(a) and (e)] effects the
transformation of the PG structure itself, accumulating more
weight in the Region (2) of Fig. 4(b) (the shoulder region would
disappear for higher fields) and Fig. 4(f). Additional
spectral-weight change, upon onset of superconductivity, on the
background of this PG is then clearly seen in Figs. 4(c)and (d)
for UD75, and Figs. 4(g) and (h) for OD88, with the growing CP
with lowering $H$ for the temperatures explored. These figures
clearly indicate that the spectral weight constituting the CP is
mainly contributed by the transfer from the bias region above the
CP position rather than from the subgap region, which contradicts
to the general perception including the BCS behavior. This is the
main finding of this work. The feature of the spectral-weight
redistribution along with the suppressed coherence in high $H$
fields is in clear contrast to previous observations in Bi-2212
\cite {mag1,mag2}, where no spectral transfer was reported either
in the subgap region \cite {mag1} or in the dip region \cite
{mag2}, despite the significantly suppressed CP by $H$ fields.

In conclusion, using our ITS, we confirmed the dominant antinodal
$c$-axis tunneling of quasiparticles as revealed in the U-shaped
spectra. Upon onset of the superconductivity in Bi-2212, the
spectral weight is transferred from the background PG hump to the
CP, generating a dip for a significant transfer at the bias
slightly above the CP position. In this sense, the dip actively
participates in the revelation of the superconductivity. This
anomalous transfer from the dip to the CP may originate from the
basic mechanism of high-$T_c$ superconductivity.

This work was supported by KOSEF through the National Research
Laboratory program. We are grateful to N. Momono, M. Oda, and M.
Ido in Hokkaido University, Japan, for providing the underdoped
single crystals for UD77 and valuable communications.


\begin{thebibliography}{00}
\bibitem{Integ} For the integration, the fluctuating zero-bias region of the curves
was smoothed out by a fourth-order polynomial fitting.
\bibitem{homogeneity} The inhomogeneous modification of
the superconducting coherence by $H$ in the presence of pancake
vortices may not be strictly equivalent to the homogeneous effect
by $T$.
\bibitem{PG structure} This statement can also be confirmed in Ref. \cite {mag2} for
the temperatures used in Fig. 4.
\end{thebibliography}
\end{document}